\documentclass [12pt]{article}
\usepackage {graphicx}
\usepackage {longtable}

\begin{document}
\begin{center}
\section*{Dirac particle in gravitational field}
\end{center}

\bigskip

\begin{center}
\textbf{M.V. Gorbatenko, T.M. Gorbatenko}
\end{center}

\bigskip

\begin{center}
Russia Federal Nuclear Center - All-Russian Scientific Research Institute of 
Experimental Physics, Sarov, Nizhny Novgorod region, E-mail: \underline 
{gorbatenko@vniief.ru}
\end{center}

\bigskip

\begin{center}
\textbf{Abstract}
\end{center}

\bigskip

Being considered is the motion of Dirac particle in gravitational field, 
described by Kerr solution. It is proved, that evolution of the wave 
function is determined by Hermitian Hamiltonian, if the concomitant 
reference frame is involved.


\subsection*{1. Introduction}


Recently a number of papers, describing the motion of Dirac particles in 
gravitational fields have been published (see, for instance [1]-[6]). The 
general logics of these papers adds up to formulating the Dirac equation in 
the form of Schrödinger equation, proving the Hermitian character of 
Hamiltonian operator, arising from the above, and analyzing the physical 
effects, resulting from the derived Hamiltonian. 

The afore described logics seems to be quite natural, however there is one 
basic obstacle to realizing the same. This obstacle is stipulated by the 
freedom of choice of the system of frame vectors, used in Dirac equation, 
influencing the physical effects mentioned. Let us go deeper into the 
essence of the problem.

As is well known, Dirac equation for bispinor $\psi $ reads as follows:

\begin{equation}
\label{eq1}
g^{\alpha \beta} \gamma _{\alpha}  \left( {\frac{{\partial \psi} }{{\partial 
x^{\beta} }} + \Phi _{\beta}  \psi}  \right) - \frac{{mc}}{{\hbar} }\psi = 
0
\end{equation}

The values $m,c,\hbar $\textit{,} constituent of (\ref{eq1}), have the regular 
meaning, that is mass, velocity of light and Plank constant, 
correspondingly. The Greek indices define vector world indices, taking on 
meanings 0,1,2,3. The signature under consideration is $\left( { - + + +}  
\right)$.

To explain the sense of value $\Phi _{\alpha}  $, constituent of equation 
(\ref{eq1}), it is necessary to introduce the system of frame vectors 
$H_{{\underline\nu}}^{\alpha}$, specified by the relation:

\begin{equation}
\label{eq2}
H_{{\underline\nu}}^{\alpha}  H_{{\underline\mu}}^{\beta}  g_{\alpha \beta}  
= \eta _{{\underline\mu}{\underline\nu}} .
\end{equation}

Where $\eta_{{\underline\mu} {\underline\nu} } $ - is a metric tensor in tangent flat 
time-space. It is assumed for the follow-up stage, that $\eta_{{\underline\mu}
{\underline\nu}} = diag\left( { - 1,1,1,1} \right)$. Matrices 
$\gamma_{{\underline\nu} } = H_{{\underline\nu}}^{\alpha}  \gamma{\alpha}$ 
will be called the frame Dirac matrices. 
They can be selected constant over the entire space. For instance, in 
standard presentation 
$\gamma_{\underline{0}} = - i\rho _{3} ;\quad \gamma_{\underline{k}} = 
\rho _{2} \sigma _{k} $. Latin indices will take on values 1,2,3. 

Further we will call $\Phi _{\alpha}  $ value as bispinor connectivity, 
which is expressed through the so called rotation Ricci coefficients 
$\Phi_{\alpha \;{\underline\mu} {\underline\nu} } $

\begin{equation}
\label{eq3}
\Phi_{\alpha \;{\underline\mu} {\underline\nu} } 
\equiv H_{{\underline\mu} }^{\sigma}  H_{{\underline\nu} \sigma 
;\alpha}  
\end{equation}

\noindent
in the following way:

\begin{equation}
\label{eq4}
\Phi_{\alpha}  = \frac{{1}}{{4}}\Phi_{\alpha \;{\underline\mu} {\underline\nu} } 
\cdot S^{{\underline\mu}{\underline\nu} }.
\end{equation}
Here

\[
H_{{\underline\nu} \sigma ;\alpha}  = H_{{\underline\nu} \sigma ,\alpha}  
-{\varepsilon\choose{\alpha\sigma}} H_{{\underline\nu}\varepsilon},
\quad
S^{{\underline\mu} {\underline\nu} } = \frac{{1}}{{2}}
\left( {\gamma^{{\underline\mu} }\gamma^{{\underline\nu} } - 
\gamma^{{\underline\nu} }\gamma^{{\underline\mu} }} \right).
\]

Frames $H_{{\underline\nu} }^{\alpha}  $ are specified with the precision up to 
Lorentz transformations, depending on coordinates. Nothing else limits the 
selection of the concrete type of frames. The associated coefficients 
$\Phi_{\alpha \;{\underline\mu} {\underline\nu}}$ correspond to each set of frames.

The problem of free choice of $H_{{\underline\nu} }^{\alpha}  $ system is solved 
differently by various authors. Most frequently this system is selected with 
consideration to the world-used coordinates and metric tensor. Assuming that 
the gravitational field is described by the metrics, considered in [1], we 
have:

\begin{equation}
\label{eq5}
ds^{2} = - V^{2}c^{2}dt^{2} + W^{2}dx^{m}dx^{n}\delta _{mn} .
\end{equation}

Here \textit{V, W -} are the two arbitrary functions of spatial coordinates. 
Then, the following system can be used as the system of frames:

\begin{equation}
\label{eq6}
H_{\underline{0}}^{\alpha}  = \frac{{1}}{{V}}\delta_{\underline{0}}^{\alpha}  ,\quad 
H_{\underline{k}}^{\alpha}  = \frac{{1}}{{W}}\delta_{\underline{k}}^{\alpha}  .
\end{equation}

In [6] the list of many other systems of frames, used by different authors, 
solving optional problems, is presented. 

Somewhat self-sufficient theory has to contain a certain principle, 
eliminating arbitrariness when selecting the frames of interest. All of the 
published up to date versions, dealing with the theory of motion of Dirac 
particles in gravitational fields, have not provided feedback to this 
challenge yet. The goal of the present paper is to formulate such a 
principle.

The approach, used for description of spin dynamics of the classic spin 
particles in gravitational field, is accepted as the starting point for 
formulating the aforementioned principle. This approach had been known at 
length (see, for instance, [7]-[8]) and is based on involving the system of 
so called concomitant frames. Under concomitant frame we will assume such a 
frame, in which the spin vector of a particle has only the spatial 
components, whereas the time spin component is equal to zero. The system of 
concomitant frames - is the combination of frames, being concomitant to each 
point of the world line of a particle.

Some time ago the above approach was used to predict and explain the 
experiment Gravity Probe B, involving gyroscopes on board of the spacecraft, 
spinning around the Earth (see [9]). In particular, the experiment recorded 
spin-orbital procession of gyroscopes, matching theoretical predictions with 
precision, better than 1\%. Thus, viability of the approach used was 
confirmed.

Dirac particle also possesses the spin, that is why implementation of 
concomitant frames for description of its spin dynamics is viewed to be 
quite appropriate. It is just the idea of concomitant frames, that is set 
forth by the authors of this paper as the only suitable method to describe 
the particle dynamics, which brings us closer to the results, recorded 
during the experiments with spin particles. 

Suitability of concomitant frame approach is demonstrated in this paper by 
the example of gravitational field, described by Kerr solution (KS). As far 
as we understand, the most essential result of our study is formulation of 
Hamiltonian, represented by Hermit operator. The results are discussed and 
are briefly compared with the results, referred from the other sources.

\subsection*{2. Dirac equation, as it is presented in reference frame 
denominations}

The following relation is true for any selected system of frames:

\begin{equation}
\label{eq7}
g^{\alpha \beta}  = - H_{\underline{0}}^{\alpha}  H_{\underline{0}}^{\beta}  
+ H_{\underline{k}}^{\alpha}H_{\underline{k}}^{\beta}  .
\end{equation}

Tensors

\begin{equation}
\label{eq8}
 - H_{\underline{0}}^{\alpha}  H_{\underline{0}}^{\beta},
\quad H_{\underline{k}}^{\alpha}H_{\underline{k}}^{\beta}  
\end{equation}

\noindent
constitute the complete system of projectors. Let us write Dirac equation 
(\ref{eq1}) so, that, the metric tensor in equation (\ref{eq1}) is exchanged for the sum of 
two projectors (\ref{eq8}).

\begin{equation}
\label{eq9}
 - H_{\underline{0}}^{\alpha}  H_{\underline{0}}^{\beta}  \gamma _{\alpha}  \left( 
{\frac{{\partial \Psi} }{{\partial x^{\beta} }} + \Phi _{\beta}  \Psi}  
\right) + H_{\underline{k}}^{\alpha}  H_{\underline{k}}^{\beta}  \gamma _{\alpha}  \left( 
{\frac{{\partial \Psi} }{{\partial x^{\beta} }} + \Phi _{\beta}  \Psi}  
\right) - \frac{{mc}}{{\hbar} }\Psi = 0.
\end{equation}

Since $H_{\underline{\mu} }^{\alpha}  \gamma_{\alpha}  = \gamma_{\underline{\mu} } $, equation 
(\ref{eq9}) is equivalent to the following:

\begin{equation}
\label{eq10}
 - \gamma_{\underline{0}} \left( {\frac{{\partial \Psi} }
{{\partial x^{\underline{0}}}} + \Phi_{\underline{0}} \Psi}  \right) + 
\gamma_{\underline{k}} \left( {\frac{{\partial \Psi 
}}{{\partial x^{\underline{k}}}} + \Phi_{\underline{k}} \Psi}  \right) - \frac{{mc}}{{\hbar 
}}\Psi = 0.
\end{equation}

The following denominations were introduced:

\begin{equation}
\label{eq11}
\frac{{\partial} }{{\partial x^{\underline{0}}}} \equiv H_{\underline{0}}^{\alpha}  
\frac{{\partial} }{{\partial x^{\alpha} }};\quad \frac{{\partial 
}}{{\partial x^{\underline{k}}}} \equiv H_{\underline{k}}^{\alpha}  \frac{{\partial} }{{\partial 
x^{\alpha} }};\quad \Phi_{\underline{0}} \equiv H_{\underline{0}}^{\alpha}  \Phi _{\alpha}  
;\quad \Phi_{\underline{k}} \equiv H_{\underline{k}}^{\alpha}  \Phi _{\alpha}  .
\end{equation}

Let us multiply equation (\ref{eq10}) on the left to 
$ - i\hbar c\gamma _{\underline{0}} $ and 
make use of identity $\gamma_{\underline{0}} \gamma_{\underline{0}} = - 1$.

\begin{equation}
\label{eq12}
 - i\hbar c\frac{{\partial \Psi} }{{\partial x^{\underline{0}}}} 
- i\hbar c\Phi_{\underline{0}} 
\Psi - i\hbar c\gamma_{\underline{0}} \gamma_{\underline{k}} 
\frac{{\partial \Psi} }{{\partial x^{\underline{k}}}} 
- i\hbar c\gamma_{\underline{0}} \gamma_{\underline{k}} \Phi_{\underline{k}} \Psi + 
imc^{2}\gamma_{\underline{0}}.
\end{equation}

In another form the above equation is written as:

\begin{equation}
\label{eq13}
i\hbar \frac{{\partial \Psi} }{{\partial \tilde {t}}} = \hat {H}\Psi ,
\end{equation}

\noindent
where $d\tilde {t} = {dx^{\underline{0}}\over{c}}$,
$\hat {H}$ has the sense of Hamilton operator,

\begin{equation}
\label{eq14}
\hat {H} = imc^{2}\gamma_{\underline{0}} + c\gamma_{\underline{0}} 
\gamma_{\underline{k}} \hat {p}_{\underline{k}} - i\hbar c\Phi_{\underline{0}} 
- i\hbar c\gamma_{\underline{0}} \gamma_{\underline{k}} \Phi_{\underline{k}} 
\end{equation}

\noindent
and operator $\hat {p}_{\underline{k}} $,constituent of $\hat {H}$, is specified by 
the relation

\begin{equation}
\label{eq15}
\hat {p}_{\underline{k}} = - i\hbar \frac{{\partial} }{{\partial x^{\underline{k}}}}.
\end{equation}

Let us call equation (\ref{eq13}) with operator $\hat {H}$ in the form (\ref{eq14}) as Dirac 
equation in reference frame denominations. As it can be seen, the structure 
of the formulated Hamiltonian is completely ``bound'' to the system of 
frames used for consideration purposes.

\subsection*{3. Frames, related to Kerr solution field}

Decompositioned components of KS metric, written in harmonic coordinates, 
were obtained in [11]. In the first nonvanishing approximation we have:

\begin{equation}
\label{eq16}
g_{\alpha \beta}  = \eta _{\alpha \beta}  + h_{\alpha \beta}  ;\quad 
g^{\alpha \beta}  = \eta ^{\alpha \beta}  - h^{\alpha \beta} 
\end{equation}

\begin{equation}
\label{eq17}
h_{00} = 2\frac{{M}}{{R}};\quad h_{0k} = 2\frac{{M\left( {J_{kl} R_{l}}  
\right)}}{{R^{3}}};\quad h_{mn} = 2\frac{{M}}{{R}}\delta _{mn} .
\end{equation}

\begin{equation}
\label{eq18}
g_{00} = - 1 + 2\frac{{M}}{{R}};\quad g_{0k} = 2\frac{{M\left( {J_{kl} 
R_{l}}  \right)}}{{R^{3}}};\quad g_{mn} = \delta_{mn} + 
2\frac{{M}}{{R}}\delta_{mn}. 
\end{equation}

The corresponding Christoffel symbols are represented in the following form:

\begin{equation}
\label{eq19}
\left. {\begin{array}{l}
{0\choose{00}}=0;\quad {0\choose{0k}}={{MR_k}\over{R^3}};\quad 
{0\choose{mn}}=3{{M\left((J_{ml}R_l)R_n +(J_{nl}R_l )R_m\right)}\over{R^5}};\\
{k\choose{00}}={{MR_k}\over{R^3}};\quad {m\choose{0n}}=2{{MJ_{mn}}\over{R^3}}-
3{{M\left(({J_{ml}R_l}R_n -({J_{nl}R_l})R_m\right)}\over{R^5}};\\
{k\choose{mn}}=-{M\over{R^3}}\left(\delta_{km}R_n +\delta_{kn}R_m -\delta_{mn}R_k\right).
\end{array}}\right\}
\end{equation}

Relations (\ref{eq2}) serve as the equations for finding the frame vectors. For the 
purposes of the present paper we will make use of solution of these 
equations in the so called symmetrical calibration. Such a solution has the 
following view:

\begin{equation}
\label{eq20}
H_{{\underline\alpha} \beta}  = \eta _{{\underline\alpha} \beta}  
+ \frac{{1}}{{2}}h_{{\underline\alpha}\beta}  .
\end{equation}

At $h_{\alpha \beta}  = 0$ the introduced in such a manner frames coincide 
with the frames of the background space, whereas at $h_{\alpha \beta}  \ne 
0$ they present the symmetrical solution of the following equation:

\[
g_{\alpha \beta}  = \eta _{\alpha \beta}  + h_{\alpha \beta}  = H_{\alpha 
}^{\underline{\mu} } \eta _{{\underline\mu} {\underline\nu} } H_{\beta} ^{{\underline\nu} } .
\]

In the first approximation for components $h_{\alpha \beta}  $ the 
difference between the world and the frame indices disappears, that is why 
omitting of underlining is admittable, in case it will not result in 
ambiguity. Having these rules at hand, we will find the components of frames 
with the subscripts.

\begin{equation}
\label{eq21}
H_{{\underline 0}0} = - \left[ {1 - \frac{{M}}{{R}}} \right];\ H_{{\underline 0}k} = 
\frac{{M\left( {J_{kl} R_{l}}  \right)}}{{R^{3}}};\ H_{{\underline k}0} = 
\frac{{M\left( {J_{kl} R_{l}}  \right)}}{{R^{3}}};\ H_{{\underline p}k} = \delta 
_{pk} \left[ {1 + \frac{{M}}{{R}}} \right].
\end{equation}

The frames with the indices of another type result from system (\ref{eq21}) with the 
corresponding indices raised up. Let us recall, that the frame indices are 
lowered or raised up using tensors $\eta_{{\underline\alpha} {\underline\beta} } ,\;\eta 
^{{\underline\alpha} {\underline\beta} }$, whereas the world indices - using tensors $g_{\alpha 
\beta}  ,\;g^{\alpha \beta} $.

\subsection*{4. The concomitant reference frames in KS field}

Let us proceed from the assumption, that the conception of mechanical 
trajectory can be introduced for Dirac particle in gravitational field, and 
that this trajectory can be described by Matisson-Papapetrou equation [12], 
[13], i.e. the following equation:

\begin{equation}
\label{eq22}
\frac{{Du^{\alpha} }}{{D\tau} } = - \frac{{1}}{{2}}\mathop {R}\limits^{\ast 
}{}^{\alpha} _{\beta \mu \nu}  u^{\beta} u^{\mu} s^{\nu} ,
\end{equation}

\noindent
where $\mathop {R}\limits^{\ast}{}^{\alpha} _{\beta \mu \nu}  $ is 
curvature tensor, dual to Riemann tensor, $\mathop {R}\limits^{\ast}  
{}^{\alpha} _{\beta \mu \nu}  \equiv E^{\alpha}_{\beta \tau \rho}  R^{\tau 
\rho}_{\mu \nu}  $. 

In KS field the part on the right in (\ref{eq22}) has the same order of smallness as 
the one on the left, consequently the motion of the particle does not add up 
to the motion over geodesic line. Coincidence objective, looking like:

\begin{equation}
\label{eq23}
\tilde {H}_{{\underline 0}}^{\alpha}  \equiv u^{\alpha} 
\end{equation}

\noindent
(the concomitant reference here and further will be marked by the upper 
tilde) means, that the concomitant frame system is non-inertial. 

Non-inertial character of $\tilde {H}_{\alpha}^{\underline{\mu} } $ system offers, 
that at each point of the world line $\tilde {H}_{\alpha}^{\underline{\mu} } $ vector 
system differs from the stationary $H_{\alpha}^{\underline{\mu} } $ vector system by 
small Lorentz transformation of booster type (see [8], [9]).

The concomitant frame will be marked by the upper tilde. Let us represent 
the combination of values $\tilde {H}_{{\alpha} \beta}  $, borrowed from [9] 
(task 11.11).

\begin{equation}
\label{eq24}
\left. {\begin{array}{l}
 {\tilde {H}_{{\underline 0}0} = \left[ { - 1 + \frac{{M}}{{R}} - \frac{{1}}{{2}}\left( 
{\dot {\eta}_{l} \dot {\eta}_{l}}  \right)} \right];\ \tilde 
{H}_{{\underline 0}k} = \dot {\eta}_{k} \left[ {1 + 3\frac{{M}}{{R}} + 
\frac{{1}}{{2}}\left( {\dot {\eta}_{l} \dot {\eta}_{l}}  \right)} \right] 
+ 2\frac{{M\left( {J_{kl} R_{l}}  \right)}}{{R^{3}}};} \\ 
 {\tilde {H}_{{\underline m}0} = - \dot {\eta} _{k} \left[ {1 + \frac{{M}}{{R}} + 
\frac{{1}}{{2}}\left( {\dot {\eta} _{l} \dot {\eta} _{l}}  \right)} 
\right];\ \tilde {H}_{{\underline m}k} = \delta _{kp} \left[ {1 + \frac{{M}}{{R}}} 
\right] + \frac{{1}}{{2}}\dot {\eta}_{k} \dot {\eta}_{p} .} \\ 
 \end{array}}  \right\}
\end{equation}

Using formulation (\ref{eq3}) we find that:

\begin{equation}
\label{eq25}
\begin{array}{l}
\tilde {\Phi}_{0}{}_{{\underline 0}{\underline k}} = \frac{{M\left( {R_{l} \dot {\eta}_{l}}  
\right)}}{{R^{3}}}\dot {\eta}_{k} - 2\frac{{M\left( {J_{kp} \dot {\eta 
}_{p}}  \right)}}{{R^{3}}} + 3\frac{{M\left[ {\left( {J_{kl} R_{l}}  
\right)\left( {R_{l} \dot {\eta}_{l}}  \right) - \left( {\dot {\eta}_{a} 
J_{ab} R_{b}}  \right)R_{k}}  \right]}}{{R^{5}}}.
\end{array}
\end{equation}

\begin{equation}
\label{eq26}
\begin{array}{l}
\tilde {\Phi}_{0}{}_{{\underline m}{\underline n}} = + \frac{{1}}{{2}}\frac{{M\left[ {R_{m} \dot 
{\eta}_{n} - R_{n} \dot {\eta}_{m}}  \right]}}{{R^{3}}} + 2\frac{{MJ_{mn} 
}}{{R^{3}}} - 3\frac{{M\left[ {\left( {J_{ml} R_{l}}  \right)R_{n} - \left( 
{J_{nl} R_{l}}  \right)R_{m}}  \right]}}{{R^{5}}}.
\end{array}
\end{equation}

 \begin{equation}
\label{eq27}
\begin{array}{l}
\tilde {\Phi}_{m}{}_{{\underline 0}{\underline k}} = \frac{{M\left( {R_{l} \dot {\eta} _{l}}  
\right)}}{{R^{3}}}\delta _{km} + 2\frac{{M}}{{R^{3}}}\dot {\eta} _{k} R_{m} 
- \frac{{M}}{{R^{3}}}\dot {\eta} _{m} R_{k}\\
- 2\frac{{MJ_{km}} }{{R^{3}}} + 
3\frac{{M\left[ {\left( {J_{kl} R_{l}}  \right)R_{m} - \left( {J_{ml} R_{l} 
} \right)R_{k}}  \right]}}{{R^{5}}}.
\end{array}
\end{equation}

\begin{equation}
\label{eq28}
\begin{array}{l}
\tilde {\Phi}{}_{p\;}{}_{{\underline m}{\underline n}} = \frac{{M\left[ {R_{m} \delta{}_{pn} - R_{n} 
\delta_{mp}}  \right]}}{{R^{3}}}.
\end{array}
\end{equation}

Bispinor connectivities $\tilde {\Phi}_{\underline{0}} $, $\tilde {\Phi}_{\underline{k}} $ can 
be found, using relations (\ref{eq4}), (\ref{eq11}).

\begin{equation}
\label{eq29}
\begin{array}{l}
\tilde {\Phi}_{\underline{0}} = \frac{{1}}{{4}}\tilde {H}_{\underline{0}}^{0} \left[ {2\tilde 
{\Phi}_{0\;{\underline 0}{\underline k}} \cdot S^{{\underline 0}{\underline k}} + \tilde {\Phi}_{0\;{\underline a}{\underline b}} \cdot 
S^{{\underline a}{\underline b}}} \right] + \frac{{1}}{{4}}\tilde {H}_{\underline{0}}^{p} \left[ {2\tilde 
{\Phi}_{p\;{\underline 0}{\underline k}} \cdot S^{{\underline 0}{\underline k}} + 
\tilde {\Phi}_{p\;{\underline a}{\underline b}} \cdot 
S^{{\underline a}{\underline b}}} \right].
\end{array}
\end{equation}

\begin{equation}
\label{eq30}
\begin{array}{l}
\tilde {\Phi}_{\underline{m}} = \frac{{1}}{{4}}\tilde {H}_{{\underline m}}^{0} \left[ {2\tilde 
{\Phi}_{0\;{\underline 0}{\underline k}} \cdot S^{{\underline 0}{\underline k}} + \tilde {\Phi}_{0\;{\underline a}{\underline b}} \cdot 
S^{{\underline a}{\underline b}}} \right] + \frac{{1}}{{4}}\tilde {H}_{{\underline m}}^{p} \left[ {2\tilde 
{\Phi}_{p\;{\underline 0}{\underline k}} \cdot S^{{\underline 0}{\underline k}} 
+ \tilde {\Phi}_{p\;{\underline a}{\underline b}} \cdot 
S^{{\underline a}{\underline b}}} \right].
\end{array}
\end{equation}

Let us substitute the expressions for frames and Ricci rotation coefficients 
into formulations for $\tilde {\Phi}_{\underline{0}} $, $\tilde {\Phi}_{\underline{k}} $. 
Having reserved the principal members of decomposition in (\ref{eq29}) and (\ref{eq30}) we 
obtain:

\begin{equation}
\label{eq31}
\tilde {\Phi}_{\underline{0}} = \frac{{1}}{{4}}\left( {\tilde {H}_{\underline{0}}^{0} 
\tilde {\Phi}_{0\;{\underline a}{\underline b}} + 
\tilde {H}_{{\underline 0}}^{p} \tilde {\Phi}_{p\;{\underline a}{\underline b}}}  
\right) \cdot S^{{\underline a}{\underline b}}.
\end{equation}

\begin{equation}
\label{eq32}
\tilde {\Phi}_{\underline{m}} = \frac{{1}}{{4}}
\tilde {H}{}_{{\underline m}}^{p} 
\tilde {\Phi}{}_{p\;{\underline a}{\underline b}} 
\cdot S^{{\underline a}{\underline b}} 
+ \frac{{1}}{{2}}\tilde {H}{}_{{\underline m}}^{p} 
\tilde {\Phi}{}_{p\;{\underline 0}{\underline k}} 
\cdot S^{{\underline 0}{\underline k}}.
\end{equation}

In KS field relations (\ref{eq31}), (\ref{eq32}) take the following form:

\begin{equation}
\label{eq33}
\begin{array}{l}
 \tilde {\Phi} {}_{\underline {0}} = \frac{{1}}{{4}}\tilde {\Phi} {}_{0\;{\underline a}{\underline b}} \cdot 
S^{{\underline a}{\underline b}} + \frac{{1}}{{4}}\dot{\eta}{}_{p} 
\tilde{\Phi} {}_{p\;{\underline a}{\underline b}} 
\cdot S^{{\underline a}{\underline b}} = \\ 
 = \frac{{3}}{{8}}\frac{{M\left[ {R_{{\underline a}} \dot {\eta} {}_{\underline{b}} 
- R_{{\underline b}} \dot{\eta} {}_{{\underline a}}}  \right]}}{{R^{3}}} 
\cdot S^{{\underline a}{\underline b}} + 
\frac{{1}}{{2}}\frac{{MJ_{{\underline a}{\underline b}}} }{{R^{3}}} 
\cdot S^{{\underline a}{\underline b}} - 
\frac{{3}}{{4}}\frac{{M\left[ {\left( {J_{{\underline a}l} R_{l}}  \right)R_{{\underline b}} 
-\left( {J_{{\underline b}l} R_{l}}  \right)R_{{\underline a}}}  \right]}}{{R^{5}}} \cdot 
S^{{\underline a}{\underline b}}. \\ 
 \end{array}
\end{equation}

\begin{equation}
\label{eq34}
\begin{array}{l}
\tilde {\Phi} {}_{\underline {m}} = \frac{{1}}{{4}}
\tilde {\Phi} {}_{{\underline m}\;{\underline a}{\underline b}} \cdot 
S^{{\underline a}{\underline b}} 
+ \frac{{1}}{{2}}\tilde {\Phi} {}_{{\underline m}\;{\underline 0}{\underline k}} 
\cdot S^{{\underline 0}{\underline k}} = 
\frac{{1}}{{4}}\frac{{M\left[ {R_{{\underline a}} \delta_{{\underline m}{\underline b}} 
- R_{{\underline b}} \delta_{{\underline m}{\underline a}}}  \right]}}{{R^{3}}} 
\cdot S^{{\underline a}{\underline b}} \\ 
+ \frac{{1}}{{2}}\left( {\frac{{M\left( {R_{\underline l} \dot {\eta}{}_{\underline l}}  
\right)}}{{R^{3}}}\delta_{\underline{km}} 
+ 2\frac{{M}}{{R^{3}}}\dot {\eta} {}_{\underline k} R_{\underline m} 
- \frac{{M}}{{R^{3}}}\dot {\eta} {}_{\underline m} R_{\underline k}}  \right) 
\cdot S^{{\underline 0}{\underline k}} \\ 
+ \frac{{1}}{{2}}\left( { - 2\frac{{MJ_{{\underline k}{\underline m}}} }{{R^{3}}} 
+ 3\frac{{M\left[ 
{\left( {J_{\underline {kl}} R_{\underline l}}  \right)R_{\underline m} 
- \left( {J_{\underline{ml}} R_{\underline l}}  \right)R_{\underline k}}  
\right]}}{{R^{5}}}} \right) \cdot S^{{\underline 0}{\underline k}}. \\ 
 \end{array}
\end{equation}

\subsection*{5. Hamiltonian}

By substituting the matrix connectivities (\ref{eq33}), (\ref{eq34}) into Hamiltonian (\ref{eq14}), 
we have:

\begin{equation}
\label{eq35}
\begin{array}{l}
\hat {H} = imc^{2}\gamma_{\underline{0}} + 
c\gamma_{\underline{0}} \gamma_{\underline{k}} \hat 
{p}{}_{\underline{k}} + i\hbar c\frac{{MR_{\underline{k}}} }{{R^{3}}} \cdot 
\gamma_{\underline{0}} \gamma_{\underline{k}} \\ 
+ 2i\hbar c\frac{{M\left( {R_{l} \dot {\eta} _{l}}  \right)}}{{R^{3}}} + 
\frac{{3}}{{8}}i\hbar c\frac{{M\left[ {R_{{\underline a}} \dot {\eta}{}_{{\underline b}} 
- R_{{\underline b}}\dot {\eta}{}_{{\underline a}}}  \right]}}{{R^{3}}} 
\cdot S^{{\underline a}{\underline b}} \\ 
+ \frac{{1}}{{2}}i\hbar c\left\{ {\frac{{MJ_{{\underline a}{\underline b}}} }{{R^{3}}} - 
\frac{{3}}{{2}}\frac{{M\left[ {\left( {J_{{\underline a}l} R_{l}}  \right)R_{\underline {b}} - 
\left( {J_{{\underline b}l} R_{l}}  \right)R_{\underline{a}}}  \right]}}{{R^{5}}}} \right\} \cdot 
S^{{\underline a}{\underline b}}. \\ 
 \end{array}
\end{equation}

Dirac equation (\ref{eq13}) with Hamiltonian (\ref{eq35}) can be written in another form:

\begin{equation}
\label{eq36}
i\hbar c\left( {\frac{{\partial} }{{\partial x^{\underline{0}}}} - 
2\frac{{M\left( {R_{l} \dot {\eta} _{l}}  \right)}}{{R^{3}}}} \right)\psi = 
\hat {\bf H}\psi .
\end{equation}

Where under operator $\hat{\bf H}$ the following operator is understood:

\begin{equation}
\label{eq37}
\begin{array}{l}
 \hat {\bf H} = imc^{2}\gamma_{\underline{0}} - i\hbar c\gamma_{\underline{0}} 
\gamma_{\underline{k}} \left( {\frac{{\partial} }{{\partial x^{\underline{k}}}}
- \frac{{MR_{\underline{k}} 
}}{{R^{3}}}} \right) \\ 
+ \frac{{3}}{{8}}i\hbar c\frac{{M\left[ {R_{\underline{a}} 
\dot {\eta}{}_{\underline{b}} - 
R_{\underline{b}} \dot {\eta} {}_{\underline{a}}}  \right]}}{{R^{3}}} 
\cdot S^{{\underline a}{\underline b}} + 
\frac{{1}}{{2}}i\hbar c\left\{ {\frac{{MJ_{{\underline a}{\underline b}}} }{{R^{3}}} 
-\frac{{3}}{{2}}\frac{{M\left[ {\left( {J_{{\underline a}l} R_{l}}  
\right)R_{\underline{b}} - 
\left( {J_{{\underline b}l} R_{l}}  \right)R_{\underline {a}}}  \right]}}{{R^{5}}}} \right\} \cdot 
S^{{\underline a}{\underline b}}. \\ 
 \end{array}
\end{equation}

\subsection*{6. Scalar product in Gilbert space}

Let us introduce Gilbert space of Dirac particle conditions for the 
concomitant reference frame. If $\phi $\textit{} and $\psi $ are two 
arbitrary vectors of condition in Gilbert space, then their scalar product 
($\phi $\textit{,$\psi $}) can be written in the form, generally accepted 
for quantum mechanics, that is in the form of integral for the 3-D space:

\begin{equation}
\label{eq38}
\left\langle {\varphi ,\psi}  \right\rangle = \int\limits_{\tilde {V}} {\bar 
{\varphi} \psi \cdot \rho d\tilde {V}} .
\end{equation}

Value $\phi \psi $\textit{}  corresponds to the product of bispinor line 
and bispinor column, and, as to its transformational properties, is a usual 
scalar complex function. The upper line over bispinor means the Dirac 
conjugation. Value $\rho $ is a so called ``weighting'' function, the form 
of which will be specified below.

Let us assume, that covariant expression for the element of the 4 volume 
$\sqrt { - g} \;d\Omega $, constructed on the four small vectors 
$dx_{\underline{0}}^{\alpha}  ,\;dx_{\underline{1}}^{\beta}  ,\;dx_{\underline{2}}^{\mu}
,\;dx_{\underline{3}}^{\nu} $, has the following view:

\begin{equation}
\label{eq39}
\sqrt { - g} \;d\Omega = \tilde {H}{}_{\alpha}^{\underline{0}} 
\tilde {H}{}_{\beta}^{\underline{1}} \tilde {H}{}_{\mu}^{\underline{2}} 
\tilde {H}{}_{\nu}^{\underline{3}} dx_{\underline{0}}^{\left[\right. 
{\alpha}}  dx_{\underline{1}}^{\beta}  dx_{\underline{2}}^{\mu}  
dx_{\underline{3}}^{{\nu}  \left.\right]} .
\end{equation}

Anti-symmetrization is provided here for the symbols, enclosed into the left 
and the right square brackets. The element of the 3 volume, for which 
integration is needed in (\ref{eq38}), has to be orthogonal to vectors $\tilde 
{H}_{\alpha} ^{{k}} $, that is why it has to be the value of the following 
view:

\begin{equation}
\label{eq40}
\sqrt { - g} \;d\Omega {}_{\alpha}^{\underline{0}} = \tilde {H}{}_{\alpha}^{\underline{0}} 
\tilde{H}{}_{\beta}^{\underline{1}} \tilde {H}{}_{\mu}^{\underline{2}} 
\tilde {H}{}_{\nu}^{\underline{3}} 
dx_{\underline{1}}^{\left[\right. {\beta}}  dx_{\underline{2}}^{\mu}  
dx_{\underline{3}}^{{\nu}  \left.\right]} .
\end{equation}

We are using the explicit expressions for the vectors of concomitant frame. 
As it turns out, in terms of the basic orders of approximation expression 
(\ref{eq40}) can be written as follows:

\begin{equation}
\label{eq41}
\begin{array}{l}
 \rho d\tilde {V} = \sqrt { - g} \;d\Omega{}_{0}^{\underline{0}} = 
\tilde {H}{}_{0}^{\underline{0}} 
\cdot det\left( {\tilde {H}{}_{m}^{\underline{k}}}  \right) 
\cdot d\tilde {V}= \\ 
= \left[ {1 - \frac{{M}}{{R}} + \frac{{1}}{{2}}\left( {\dot {\eta} {}_{l} 
\dot {\eta} {}_{l}}  \right)} \right]\left[ {1 + 3\frac{{M}}{{R}} + 
\frac{{1}}{{2}}\left( {\dot {\eta} {}_{l} \dot {\eta} {}_{l}}  \right)} 
\right]d\tilde {V} = \left[ {1 + 2\frac{{M}}{{R}} + \left( {\dot {\eta} _{l} 
\dot {\eta} _{l}}  \right)} \right]d\tilde {V}. \\ 
\end{array}
\end{equation}

From the above we obtain the expression for the ``weighting'' function:

\begin{equation}
\label{eq42}
\rho = \left[ {1 + 2\frac{{M}}{{R}} + \left( {\dot {\eta} _{l} \dot {\eta 
}_{l}}  \right)} \right].
\end{equation}

\subsection*{7. Hermit character of Hamiltonian}

Assume, that some operator  is specified in state space. If for any two 
vectors of condition $\phi $\textit{} and $\psi $\textit{}  the following 
relation is true:

\begin{equation}
\label{eq43}
\left\langle {\varphi ,\hat {A}\psi}  \right\rangle = \left\langle {\hat 
{A}\varphi ,\psi}  \right\rangle ,
\end{equation}

\noindent
then, operator  can be called the Hermit one. Hermit property will be 
expressed as $\hat {A}^{ +}  = \hat {A}$.

Presence of the ``weighting'' multiplier $\rho $ in expression (\ref{eq38}) does not 
effect the Hermit properties of those parts of operators, which do not 
contain differentiation operations. In case the operator contains 
differentiation operation, identification of Hermit properties requires 
special consideration. For instance, pulse operator enters Hamiltonian in 
the following combination:

\begin{equation}
\label{eq44}
\left\{ {c\gamma_{\underline{0}} \gamma_{\underline{k}} \hat{p}{}_{\underline{k}} 
+ i\hbar c\frac{{MR_{\underline{k}}} }{{R^{3}}} \cdot \gamma_{\underline{0}} 
\gamma_{\underline{k}}}  \right\}.
\end{equation}

For such a combination we have:

$$
\begin{array}{l}
\left\langle {\left( {c\gamma_{\underline{0}} \gamma_{\underline{k}} 
\hat {p}{}_{\underline{k}} + i\hbar 
c\frac{{MR_{\underline{k}}} }{{R^{3}}} \cdot \gamma_{\underline{0}} 
\gamma_{\underline{k}}}  
\right)\varphi ,\psi}  \right\rangle=\\ 
= \left\langle {\left( {c\gamma_{\underline{0}} 
\gamma_{\underline{k}} \hat {p}{}_{\underline{k}}}  \right)\varphi ,\psi}  
\right\rangle + 
\left\langle {\left( {i\hbar c\frac{{MR_{\underline{k}}} }{{R^{3}}} 
\cdot \gamma_{\underline{0}} \gamma_{\underline{k}}}  
\right)\varphi ,\psi}  \right\rangle. 
\end{array}
$$

After substituting the explicit expression (\ref{eq15}) for operator $\hat {p}_{{k}} 
$ and integration by parts we obtain the following:

\[
\begin{array}{l}
 \left\langle {\left( {c\gamma_{\underline{0}} \gamma_{\underline{k}} 
\hat {p}{}_{\underline{k}} + i\hbar 
c\frac{{MR_{\underline{k}}} }{{R^{3}}} \cdot \gamma_{\underline{0}} 
\gamma_{\underline{k}}}  
\right)\varphi ,\psi}  \right\rangle =\\
= i\hbar c\int\limits_{V} 
{\frac{{\partial \varphi{}^{ +} }}{{\partial x^{\underline{k}}}}
\gamma_{\underline{0}} \gamma_{\underline{k}} \psi \rho d\tilde {V}} 
- i\hbar c\int\limits_{V} {\varphi^{ +}\gamma_{\underline{0}} 
\gamma_{\underline{k}} \psi \rho dV} =\\ 
 = - i\hbar c\int\limits_{V} {\varphi ^{ +} \gamma_{\underline{0}} 
\gamma_{\underline{k}} 
\frac{{\partial \psi} }{{\partial x^{\underline{k}}}}\rho d\tilde {V}} - i\hbar 
c\int\limits_{V} {\varphi^{ +} \gamma_{\underline{0}} \gamma_{\underline{k}} 
\psi \frac{{\partial \rho} }{{\partial x^{\underline{k}}}}d\tilde {V}} - i\hbar 
c\int\limits_{V} {\varphi ^{ +} \gamma_{\underline{0}} \gamma_{\underline{k}} 
\psi \rho dV} . \\ 
 \end{array}
\]

After substituting expression (\ref{eq42}) for the ``weighting'' function $\rho $ we 
have the following:

\[
\begin{array}{l}
 \left\langle {\left( {c\gamma_{\underline{0}} \gamma_{\underline{k}} 
\hat {p}{}_{\underline{k}} + i\hbar 
c\frac{{MR_{\underline{k}}} }{{R^{3}}} \cdot \gamma_{\underline{0}} 
\gamma_{\underline{k}}}  
\right)\varphi ,\psi}  \right\rangle  
 = \left\langle {\varphi ,\left( {c\gamma_{\underline{0}} 
\gamma_{\underline{k}} \hat{p}{}_{\underline{k}}}  \right)\psi}  
\right\rangle +\\
+ 2i\hbar c\int\limits_{V} {\varphi^{ +} \gamma_{\underline{0}} 
\gamma_{\underline{k}} \psi \left( {\frac{{MR_{\underline k}} }{{R^{3}}}} 
\right)dV} - i\hbar c\int\limits_{V} {\varphi^{ +} \gamma_{\underline{0}} 
\gamma_{\underline{k}} \psi \left( {\frac{{MR_{\underline k}} }{{R^{3}}}} 
\right)dV} = \\ 
= \left\langle {\varphi ,\left( {c\gamma_{\underline{0}} 
\gamma_{\underline{k}} \hat{p}{}_{\underline{k}} 
+ i\hbar c\frac{{MR_{{k}}} }{{R^{3}}} \cdot \gamma_{\underline{0}} 
\gamma_{\underline{k}}}  \right)\psi}  \right\rangle . \\ 
 \end{array}
\]

From the above it follows, that:

\begin{equation}
\label{eq45}
\left( {c\gamma_{\underline{0}} \gamma_{\underline{k}} 
\hat {p}{}_{\underline{k}} + i\hbar 
c\frac{{MR_{\underline{k}}} }{{R^{3}}} \cdot \gamma_{\underline{0}} 
\gamma_{\underline{k}}}  \right)^{ + 
} = \left( {c\gamma _{{0}} \gamma _{{k}} \hat {p}_{\underline{k}} 
+ i\hbar c\frac{{MR_{\underline{k}}} }{{R^{3}}} \cdot \gamma_{\underline{0}} 
\gamma_{\underline{k}}}  \right).
\end{equation}

With consideration of the above relation we may come to the conclusion, that 
Hamiltonian (\ref{eq37}) is indeed the Hermit operator.

Analogously it is being proved, that the operator on the left part of 
equation (\ref{eq36}) is also the Hermit operator, that is what is being proved is:

\begin{equation}
\label{eq46}
\left[ {i\hbar c\left( {\frac{{\partial} }{{\partial x^{\underline{0}}}} - 
2\frac{{M\left( {R_{l} \dot {\eta} {}_{l}}  \right)}}{{R^{3}}}} \right)} 
\right]^{ +}  = i\hbar c\left( {\frac{{\partial} }{{\partial x^{\underline{0}}}} 
- 2\frac{{M\left( {R_{l} \dot {\eta} {}_{l}}  \right)}}{{R^{3}}}} \right).
\end{equation}

Thus, Hermit operators are present in both parts of equation (\ref{eq36}).

\subsection*{8. Discussion of the results}

The present paper considers, within the concept of ``test particles'', the 
dynamics of Dirac particles in gravitational field, generated by a massive 
rotating body. The levels of freedom, describing the motion, are separated 
into two groups for the purposes of the above consideration effort. Those, 
defining the trajectory of particle motion, are found quasi-classically, 
using Matisson-Papapetrou equation (\ref{eq22}). The levels of freedom, related to 
the wave function, are found by solving Dirac equation in the concomitant 
frame system (\ref{eq24}), defined by the world line of the particle and the 
parameters of gravitational field.

All the members, constituent of the Hamiltonian under consideration (\ref{eq37}), 
have the direct physical sense. Undergoing changes here are only pulse 
operator $ - i\hbar \frac{{\partial} }{{\partial x^{\underline{k}}}}$ and energy 
operator $i\hbar \frac{{\partial} }{{\partial \tilde {t}}}$. In 
gravitational field the analogous to the above are the following operators:

\[
 - i\hbar \left( {\frac{{\partial} }{{\partial x^{\underline{k}}}} 
- \frac{{MR_{\underline{k}} 
}}{{R^{3}}}} \right),\quad i\hbar c\left( {\frac{{\partial} }{{\partial 
x^{\underline{0}}}} - 2\frac{{M\left( {R_{l} \dot {\eta}{}_{l}}  \right)}}{{R^{3}}}} 
\right).
\]

With consideration to scalar product in Hilbert space specified, and the 
``weighting'' function (\ref{eq42}), which occurred in the course of our 
consideration, we may conclude, that the above operators possess Hermit 
property. As we envision, occurrence of the ``weighting'' function of $\rho 
$\textit{} type (\ref{eq42}) within our study is a non-trivial circumstance. For 
instance, using stationary system of references $H_{\alpha}^{\underline{\mu} } $, 
related to KS field, does not result, as our calculations prove, in 
occurrence of the applicable ``weighting'' function. 

Our study demonstrates, that implementation of the concomitant system of 
frames $\tilde {H}{}_{\alpha}^{\underline{\mu} } $ in KS gravitational field allows 
writing Dirac equation in the form of equation (\ref{eq36}), in which operators in 
the left and the right parts are of the Hermit character.

When finding spin evolution of Dirac particle, the members of two types, 
depending on the spin effect, will enter Hamiltonian $\hat {\bf H}$ (\ref{eq37}). The 
first member type will depend on the particle velocity and will be 
describing the geodesic precession of spin. The second member type has the 
spin-spin structure and describes Lense-Thirring precession. As a result we 
may conclude, that in the approximation under study the spin behavior of 
Dirac particle is analogous to the spin behavior of classical spin 
particles. 

It follows from the above, that considerations, concerning the character of 
equivalence principle, as it is performed in part, dealing with spin motion, 
remain valid for Dirac particles. This principle shows up as non-dependence 
of frequency of precessions from the spin value: the Dirac particle with 
${{\hbar}  \mathord{\left/ {\vphantom {{\hbar}  {2}}} \right. 
\kern-\nulldelimiterspace} {2}}$ spin precesses in the same way as the 
classical spin particles. At the same time the world lines of spin particles 
in KS field do not coincide neither with each other, nor with the geodesic 
ones, which is the evidence of the fact, that the equivalence principle does 
not work here.

It is easy to make sure, that all the above results transform naturally into 
the results, corresponding to Schwarzschild solution, if $J_{mn}$ is 
assumed to equal to zero, that is the spin tensor of the ``big body'', 
giving rise to gravitational field, is also assumed to be equal to zero. 

Let us pay attention to one more aspect of the system of concomitant frames. 
What is meant is defining the intrinsic time of the particle $d\tau = - 
\frac{{1}}{{c}}\left( {u_{\nu}  dx^{\nu} } \right)$, as it moves over the 
proper world line. If the congruence of coordinate curves is chosen, one of 
which coincides with the world line of the particle, then value $d\tau $ can 
be found from relation $d\tau = \sqrt { - g_{00}}  \;dt$, when the particle 
drifts for $dx^{\alpha} $ over the motion trajectory. By integrating $d\tau 
$ along the world line we will obtain the ``true'' time of the particle, 
i.e. the time, which would have been demonstrated by the atomic clock, 
should it be resting relative to the particle. Such a method of time 
specification coincides with the method, applied in General Relativity 
Theory, when calculating the intrinsic time of the observer, moving randomly 
(see, for instance, [13], \S{84}). It can be referred from the above, that 
implementing the concomitant system of frames essentially means implementing 
such a reference system, for which the laws of quantum mechanics are true, 
defining, in particular, the rate of the atomic clock. It is not 
accidentally, that just in such a system the Dirac equation can be written 
in the form of Schrödinger equation with Hermit Hamiltonian. 

According to our opinion, those proceedings, in which $\tilde {H}_{\alpha 
}^{{\mu} } $ system is not involved, but the Hermit character of Hamiltonian 
is still achieved, implement some ``illegal'' operation somewhere. The 
transformation, presented in paper [1] may serve as an example of such an 
operation:

\[
\psi \to {\psi} ' = W^{3/2}\psi ,\quad \quad \hat {H} \to {\hat {H}}' = 
W^{3/2}\hat {H}W^{ - 3/2}
\]

\noindent
which transformation is not unitary.

\subsection*{Acknowledgments}

One of the authors (M.G.) expresses his deep appreciation to V.P. Neznamov 
for a number of fruitful discussions of the problems, being the target of 
the present paper.

\subsection*{References}

[1] Yu.N.Obukhov. \textit{Spin, gravity, and inertia.} ArXiv: gr-qc/0012102.

[2] Silenko Alexander J., Teryaev Oleg V. \textit{Semiclassical limit for 
Dirac particles interacting with a gravitational field}// Phys. Rev. 
D\textbf{71}, 064016 (2005). 

[3] Arminjon Mayeul. \textit{Post-Newtonian equation for the energy levels 
of a Dirac particle in a static metric}// Phys. Rev. D\textbf{74}, 065017 
(2006).

[4] A.A.Pomeransky, R.A.Sen'kov, I.B.Khriplovich.\textit{Spinning relativistic 
particles in external fields}// UFN. \textbf{170}, No.10. P.1129 (2000).

[5] I.B.Khriplovich. \textit{Spinning Relativistic Particles in External 
Fields}. E-print. arXiv: 0801.1881v1 [gr-qc].

[6] A.J.Silenko. \textit{Classical and quantum spins in curved spacetimes}. 
E-print. arXiv: 0802.4443v1 [gr-qc].

[7] S.Weinberg. \textit{Gravitation and Cosmology}. John Wiley \& Sons. 
Hoboken. (1972).

[8] C.W.Misner, K.S.Thorne, and J.A.Wheeler. \textit{Gravitation}. 
W.H.Freeman, San Francisco (1973).

[9] A.P.Lightman, W.H.Press, R.H.Price, and S.A.Teukolsky. \textit{Problem 
book in relativity and gravitation.} Princeton Univ. Press 1975.

[10] \underline{www.einstein.stanford.edu}. Final Report 
on the experiment Gravity Probe B.

[11] M.V.Gorbatenko, T.M.Gorbatenko// Theor. and Math. Phys. \textbf{140}, 
Issue 1, July 2004, P. 1028.

[12] M.Mathisson. Acta Phys. Pol. \textbf{6}. 163 (1937).

[13] A.Papapetrou. Proc. Roy. Soc. Lond. \textbf{A209}. 248 (1951).

[14] L.D.Landau, E.M.Lifshits. \textit{The field theory}. Moscow. Nauka 
Publishers (1988).

\end{document}